\documentclass{llncs}

\usepackage{cite}
\usepackage{graphicx}
\usepackage{amsfonts,amssymb,amsmath,alltt}
\usepackage{array}
\usepackage{mdwmath}
\usepackage{mdwtab}
\usepackage{fixltx2e}
\usepackage{stfloats}
\usepackage{hyperref}
\usepackage{flushend}
\usepackage{xspace}
\usepackage{color}
\usepackage{stmaryrd}

\newenvironment{ttbox}{\begin{alltt}\ttbraces\small\tt}%
                      {\end{alltt}}
\def\ttbraces{\let\.=\nobreak\chardef\{=`\{\chardef\}=`\}\chardef\|=`\\}

\newcommand\ttand{\mbox{{$\land$}}}
\newcommand\ttor{\mbox{{$\lor$}}}
\newcommand\ttfun{\mbox{{$\Rightarrow$}}}

\newcommand\ttcup{\mbox{{$\cup$}}}

\newcommand\ttimp{\mbox{{$\longrightarrow$}}}
\newcommand\ttequiv{\mbox{{$\equiv$}}}
\newcommand\ttexists{\mbox{{$\exists$}}}
\newcommand\ttforall{\mbox{{$\forall$}}}
\newcommand\ttneg{\mbox{{$\neg$}}}
\newcommand\ttneq{\mbox{{$\neq$}}}
\newcommand\ttin{\mbox{{$\in$}}}
\newcommand\ttnin{\mbox{{$\notin$}}}
\newcommand\ttImp{\mbox{{$\Longrightarrow$}}}
\newcommand\ttlam{\mbox{\( \lambda \)}}
\newcommand\tttimes{\mbox{\( \times \)}}
\newcommand\ttlbrack{\mbox{\(\llbracket\)}}
\newcommand\ttrbrack{\mbox{\( \rrbracket \)}}

\newcommand\ttatI{\mbox{\( @_G \)}}

\newcommand\ttrelI{\mbox{{$\to_{i}$}}}

\newcommand\ttsigma{\mbox{{$\sigma$}}}

\newcommand\ttsubseteq{\mbox{{$\subseteq$}}}
\newcommand\ttsupseteq{\mbox{{$\supseteq$}}}
\newcommand\ttf{\mbox{{$f$}}}
\newcommand\ttvdash{\mbox{{$\vdash$}}}
\newcommand\ttvdashV{\mbox{{$\vdash_V$}}}
\newcommand\ttref{\mbox{{$\sqsubseteq$}}}
\newcommand\ttrefV{\mbox{{$\sqsubseteq_V$}}}
\newcommand{\ttcalN}[1]{\mbox{{${\cal{N}}_{\texttt{#1}}$}}} 
\newcommand\ttattand[1]{\mbox{{$\oplus_{\wedge}^{#1}$}}}
\newcommand\ttattor[1]{\mbox{{$\oplus_{\vee}^{#1}$}}}
\newcommand\ttrelIstar{\mbox{{$\to_{i}^*$}}}
\newcommand\ttrel[1]{\mbox{{$\to_{#1}$}}}

\newcommand\ttupdownarrow{\mbox{{$\Updownarrow$}}}

\begin{document}
\frontmatter
  
\mainmatter

\title{Attack Trees in Isabelle 
}
\author{Florian Kamm\"uller}

\institute{Middlesex University London and\\ Technische Universit\"at Berlin\\
\email{f.kammueller@mdx.ac.uk}
}
\maketitle

\begin{abstract}
In this paper, we present a proof theory for attack trees. Attack trees are a
well established and useful model for the construction of attacks on systems
since they allow a stepwise exploration of high level attacks in 
application scenarios. Using the expressiveness of Higher Order Logic in 
Isabelle, we succeed in developing a generic theory of attack trees with 
a state-based semantics based on Kripke structures and CTL. The resulting 
framework allows mechanically supported logic analysis of the meta-theory
of the proof calculus of attack trees and at the same time the developed 
proof theory enables application to case studies. A central correctness 
and completeness result proved in  Isabelle establishes a connection 
between the notion of attack tree validity and CTL. 
The application is illustrated on the example of a healthcare IoT system
and GDPR compliance verification.
\end{abstract}

\section{Introduction}
Attack trees are an intuitive and practical formal method to analyse and quantify
attacks on security and privacy. They are very useful to identify the steps an attacker
takes through a system when approaching the attack goal. In this paper, we provide 
a proof calculus to analyse concrete attacks using a notion of attack validity.
We define a state based semantics with Kripke models and the temporal logic
CTL in the proof assistant Isabelle \cite{npw:02} using its Higher Order Logic 
(HOL)\footnote{In the following, we refer to Isabelle/HOL simply as Isabelle.}. 
We prove the correctness and completeness (adequacy) of attack trees in Isabelle 
with respect to the model.
This generic Kripke model enriched with CTL does not use an action based model 
contrary to the main stream. Instead, our model of attack trees leaves the choice 
of the actor and action model to the application.
Nevertheless, using the genericity of Isabelle, proofs and concepts of attack
trees carry over to the application model.

There are many approaches to provide a mathematical and formal semantics as
well as constructing verification tools for attack trees but we pioneer 
the use of a Higher Order Logic (HOL) tool like Isabelle that allows
proof of meta-theory -- like adequacy 
of the semantics -- and verification of applications -- while being ensured 
that the formalism is correct.

Attack trees have been investigated on a theoretical level quite intensively;
various extensions exist, e.g., to attack-defense trees and probabilistic or 
timed attack trees. This paper uses preliminary work towards an
Isabelle proof calculus for attack trees presented at a workshop 
\cite{kam:17a} but accomplishes the theoretical foundation by
defining a formal semantics and providing the proof of correctness and 
completeness and thereby establishing a feasible link for application 
verification.
The novelty of this proof theoretic approach to attack tree verification
is to take a logical approach from the very beginning by imposing the 
rigorous expressive Isabelle framework as the technical and semantical 
spine. This approach brings about a decisive advantage which is beneficial 
for a successful application of the attack tree formalism and consequently 
also characterizes our contribution: meta-theory and application verification 
are possible simultaneously.
Since Higher Order Logic allows expressing concepts like attack trees 
within the logic, it enables reasoning {\it about} objects like attack 
trees, Kripke structures, or the temporal logic CTL in the logic 
(meta-theory) while  at the same time {\it applying} these formalised concepts 
to applications like infrastructures with actors and policies (object-logics).

This paper presents the following contributions.
\begin{itemize}
\item We provide a proof calculus for attack trees that entails a
notion of refinement of attack trees and a notion of valid attack trees.
\item Validity of attack trees can be characterized by a recursive 
function in Isabelle which facilitates evaluation and permits code 
generation.
\item The main theorems show the correctness and completeness of attack tree 
validity with respect to the state transition semantics based on Kripke structures
and CTL. This meta-theorem not only provides a proof for the concepts
but is part of the proof calculus for applications.
\item The Isabelle attack tree formalisation is applied to the case study
of formalising GDPR properties over infrastructures.
\end{itemize}

In this paper, we first introduce the underlying Kripke structures and CTL
logic (Section \ref{sec:kripkectl}). Next, we present attack trees and
their notion of refinement (Section \ref{sec:atref}). The notion of validity 
is given by the proof calculus in Section \ref{sec:procal} followed
by the central theorem of correctness and completeness (adequacy)
of attacks in Section \ref{sec:cor}
including a high level description of the proof. Section \ref{sec:iothc}
shows how the framework is applied to analyse an IoT healthcare system
and Section \ref{sec:gdpr} extends by labelled data to enable 
GDPR compliance verification.
We then discuss, consider related work, and draw conclusions (Section \ref{sec:concl}).
All Isabelle sources are available online \cite{kam:18smc}.

\section{Kripke Structures and CTL in Isabelle}
\label{sec:kripkectl}
Isabelle is a generic Higher Order Logic (HOL) proof assistant. Its generic
aspect allows the embedding of so-called object-logics as new theories
on top of HOL. There are sophisticated proof tactics available to support 
reasoning: simplification, first-order resolution, and special macros to support
arithmetic amongst others.
The use of HOL has the advantage that it enables expressing
even the most complex application scenarios, conditions, and logical
requirements and HOL simultaneously enables the analysis of the meta-theory. 

Object-logics, when added to Isabelle using constant and type definitions,
constitute a so-called {\it conservative extension}. This means that no 
inconsistency can be introduced; conceptually, new types are defined as 
subsets of existing types and properties are proved using a one-to-one 
relationship to the new type from properties of the existing type.

\begin{figure}[h!]
\begin{center}
\includegraphics[scale=.4]{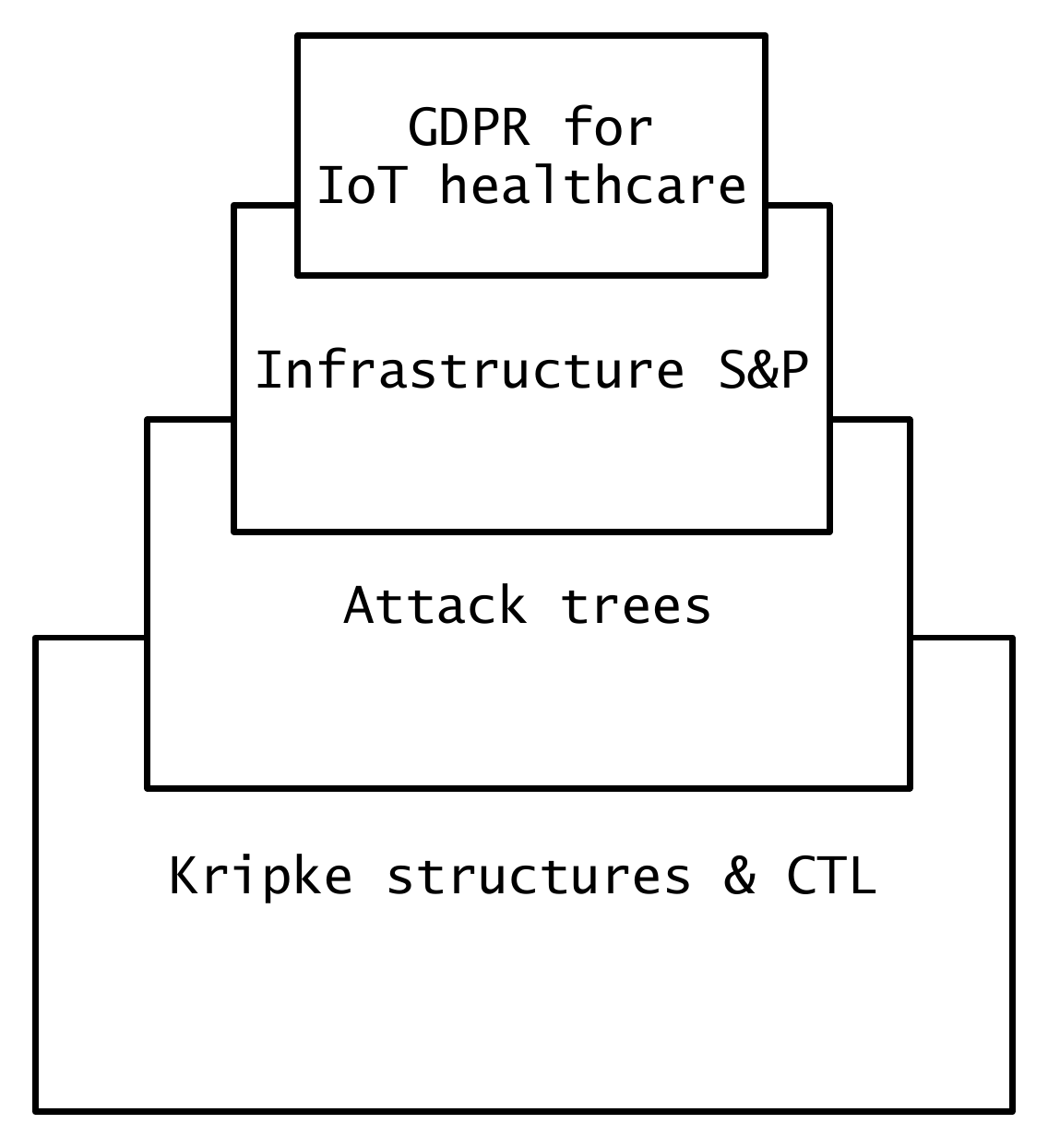}
\end{center}
\vspace{-.5cm}
\caption{Generic framework for attack trees embeds applications.}
\label{fig:theorystruc}
\end{figure}

In this work, we make additional use of the class concept of Isabelle that 
allows an abstract specification of a set of types and properties to be 
instantiated later. We use it to abstract from states and state transition
in order to create a generic framework for
Kripke structures, CTL, and attack trees. Using classes the framework can 
then be applied to arbitrary object-logics that have a notion of state and
state transition by instantiation.
Isabelle attack trees have been designed as a generic framework meaning that 
the formalised theories can be applied to various applications.
Figure \ref{fig:theorystruc} illustrates how the Isabelle theories in our
framework are embedded into each other.

\subsection{Kripke Structures and CTL}
\label{sec:kripke}
We apply Kripke structures and CTL to model state based systems and analyse
properties under dynamic state changes.
Snapshots of systems are the states on which we define a 
state transition. Temporal logic is then employed to express security and privacy
properties.  

In Isabelle, the system states and their transition relation are defined as a 
class called \texttt{state} containing an abstract constant \texttt{state\_transition}. 
It introduces the syntactic infix notation \texttt{I \ttrelI\, I'} to denote 
that system state \texttt{I} and \texttt{I'} are in this relation over an arbitrary 
(polymorphic) type $\ttsigma$.
\begin{ttbox}
{\bf class} state = 
{\bf fixes} state_transition :: [\ttsigma :: type, \ttsigma] \ttfun bool ("_  \ttrelI _")
\end{ttbox}
The above class definition lifts Kripke structures and CTL to 
a general level.
The definition of the inductive relation is given by a set of specific rules
which are, however, 
part of an application like infrastructures (Section \ref{sec:iothc}).
Branching time temporal logic CTL 
is defined in general over Kripke structures with arbitrary state transitions 
and can later be applied to suitable theories, like infrastructures. 

Based on the generic state transition $\ttrelI$ of the type class \texttt{state},
the CTL-operators \texttt{\sf EX} and \texttt{\sf AX} express that property $f$ 
holds in some or all next states, respectively.
The CTL formula \texttt{\sf AG} $f$ means that on all paths branching from 
a state $s$ the formula $f$ is always true (\texttt{\sf G} stands for `globally'). 
It can be defined using the Tarski fixpoint theory by applying the greatest 
fixpoint operator.
In a similar way, the other CTL operators are defined. 
The formal Isabelle definition of what it means that formula 
$f$ holds in a Kripke structure \texttt{M} can be
stated as: the initial states of the Kripke structure \texttt{init M} 
need to be contained in the set of all states \texttt{states M} 
that imply $f$.
\begin{ttbox}
 M \ttvdash \ttf \ttequiv  init M \ttsubseteq \{ s \ttin states M. s \ttin \ttf  \}
\end{ttbox}
In an application, the set of states of the Kripke structure will be defined 
as the set of states reachable by the infrastructure state transition from 
some initial state, say \texttt{example\_scenario}.
\begin{ttbox}
  example\_states \ttequiv \{ I. example\_scenario \ttrelI\^{}*  I \}
\end{ttbox}
The relation \texttt{\ttrelI\^{}*} is the reflexive transitive closure -- an operator
supplied by the Isabelle theory library -- applied to the relation \texttt{\ttrelI}.

The \texttt{Kripke} constructor combines the constituents initial state, state set and
state transition relation \texttt{\ttrelI}. 
\begin{ttbox}
 example_Kripke \ttequiv Kripke example_states \{example_scenario\} \ttrelI 
\end{ttbox}
In Isabelle, the concept of sets and predicates coincide\footnote{In general,
this is often referred to as {\it predicate transformer semantics.}}.
Thus a \texttt{property} is a predicate over states which is equal to a set of 
states. For example, we can then try to prove that there is a path ({\sf E}) 
to a state in which the property eventually holds (in the {\sf F}uture) by 
starting the following proof in Isabelle.
\begin{ttbox}
  example_Kripke \ttvdash {\sf EF} property 
\end{ttbox}

\section{Attack Trees and Refinement}
\label{sec:atref}
\begin{figure}[h!]
\begin{center}
\includegraphics[scale=.25]{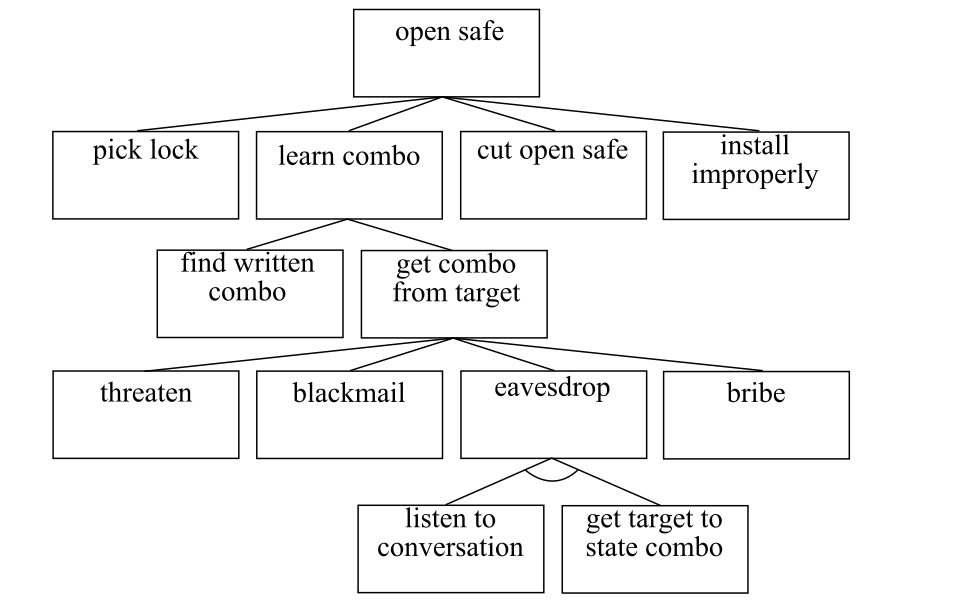}
\end{center}
\caption{Attack tree example illustrating disjunctive nodes for alternative 
attacks refining the attack ``open safe''. Near the leaves there is also a 
conjunctive node ``eavesdrop''.}
\label{fig:atex}
\end{figure}

Attack Trees \cite{Schneier.102} are a graphical language for the analysis 
and quantification of attacks. If the root represents an attack, 
its children represent the sub-attacks. 
Leaf nodes are the basic attacks; other
nodes of attack trees represent sub-attacks.
Sub-attacks can be alternatives for reaching the goal (disjunctive node) or 
they must all be completed to reach the goal (conjunctive node). 
Figure \ref{fig:atex} is an example of an attack tree taken from a textbook
\cite{Schneier.102} illustrating the attack of opening a safe.
Nodes can be adorned with attributes, for example costs of attacks
or probabilities which allows quantification of attacks (not used 
in the example).

\subsection{Attack Tree Datatype in Isabelle}
\label{sec:data}
The following datatype definition \texttt{attree} defines attack trees.
The simplest case of an attack tree is a base attack.
The principal idea is that base attacks are defined by a pair of
state sets representing the initial states and the {\it attack property}
-- a set of states characterized by the fact that this property holds
in them. 
Attacks can also be combined as the conjunction or disjunction of other attacks. 
The operator $\oplus_\vee$ creates or-trees and $\oplus_\wedge$ creates and-trees.
And-attack trees $l \ttattand s$ and or-attack trees $l \ttattor s$ 
consist of a list of sub-attacks -- again attack trees. 
\begin{ttbox}
{\bf datatype} (\ttsigma :: state)attree = 
  BaseAttack (\ttsigma set)\tttimes(\ttsigma set) ("\ttcalN (_)") 
| AndAttack (\ttsigma attree)list (\ttsigma set)\tttimes(\ttsigma set) ("_ {\ttattand{(_)}}")
| OrAttack  (\ttsigma attree)list (\ttsigma set)\tttimes(\ttsigma set) ("_ {\ttattor{(_)}}")
\end{ttbox}
The attack goal 
is given by the pair of state sets on the right of the operator 
\texttt{\ttcalN}, $\oplus_\vee$ or $\oplus_\wedge$, respectively. A corresponding 
projection operator is defined as the function \texttt{attack}.
\begin{ttbox}
{\bf primrec} attack :: (\ttsigma::state)attree \ttfun (\ttsigma set)\tttimes(\ttsigma set)
{\bf where} 
  attack (BaseAttack b) = b 
| attack (AndAttack as s) = s 
| attack (OrAttack as s) = s
\end{ttbox}


\subsection{Attack Tree Refinement}
\begin{figure*}
\begin{center}
\includegraphics[scale=.3]{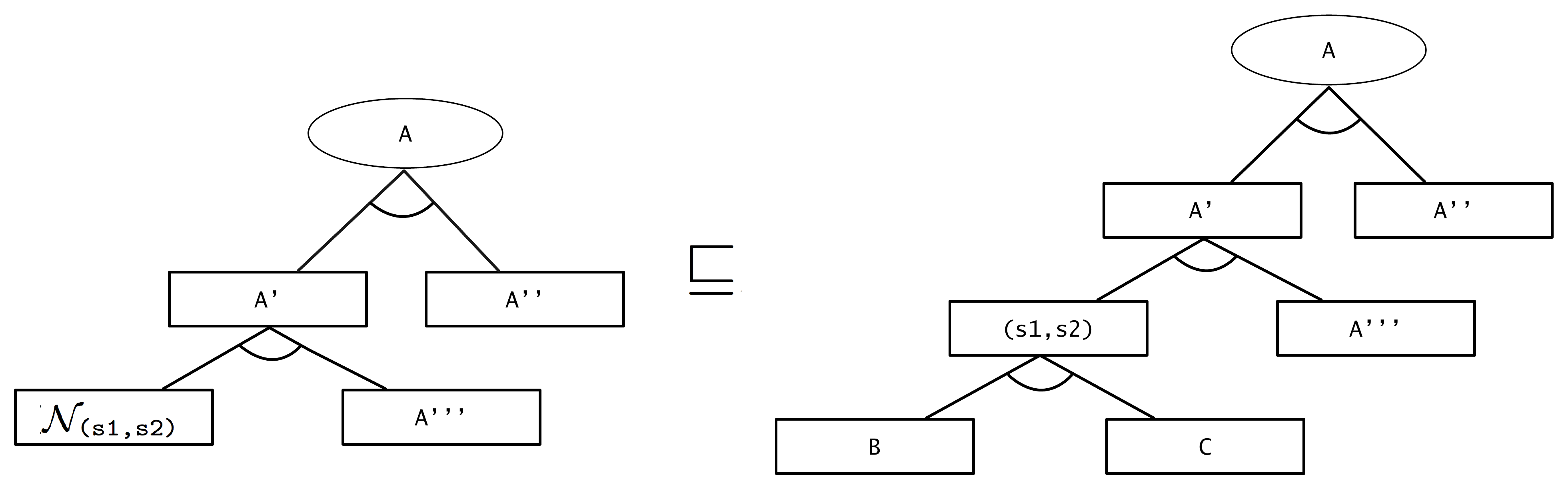}
\end{center}
\caption{Attack tree example illustrating refinement of an and-subtree.}
\label{fig:ref}
\end{figure*}

When we develop an attack tree, we proceed from an abstract attack, given
by an attack goal, by breaking it down into a series of sub-attacks. This
proceeding corresponds to a process of {\it refinement}. Therefore, as part of
the attack tree calculus, we provide a notion of attack tree refinement.
This can be done elegantly by defining an infix operator $\sqsubseteq$. 
The intuition of developing an attack tree from the root to the leaves
is illustrated  in Figure \ref{fig:ref}. The example attack tree
on the left side has a leaf that is expanded by the refinement 
into an and-attack with two steps.
Formally, we define the semantics of the refinement operator by an
inductive definition for the constant \texttt{\ttref}, that is, the 
smallest predicate closed under the set of specified rules.
\begin{ttbox}
{\bf inductive} refines_to :: [(\ttsigma :: state) attree, \ttsigma attree] \ttfun bool ("_ \ttref _")
{\bf where} 
  refI: \ttlbrack  A = (l @ [\ttcalN{(s1,s2)}] @ l'')\ttattand{(s0,s3)}; A' = l' \ttattand{(s1,s2)};
          A'' = l @ l' @ l'' \ttattand{(s0,s3)} \ttrbrack \ttImp A \ttref A'' 
| ref_or: \ttlbrack as \ttneq []; \ttforall A' \ttin set(as). A \ttref A' \ttand attack A = s 
          \ttrbrack \ttImp A \ttref as \ttattor s 
| ref_trans: \ttlbrack A \ttref A'; A' \ttref A'' \ttrbrack \ttImp A \ttref A''
| ref_refl : A \ttref A
\end{ttbox}
The rule \texttt{refI} captures the intuition expressed in Figure \ref{fig:ref}:
a sequence of leaves in an and-subtree can be refined by
replacing a single leaf by a new subsequence (the \texttt{@} is the 
list append in Isabelle).
Rule \texttt{ref\_or} describes or-attack refinement. To refine a node into 
an or-attack, all sub-trees in the or-attack list need to refine the parent 
node. 
The remaining rules define \texttt{\ttref}
as a pre-order on sub-trees of an attack tree: it is reflexive and transitive. 

Refinement of attack trees defines the stepwise process of expanding abstract
attacks into more elaborate attacks only syntactically. There is no guarantee
that the refined attack is possible if the abstract one is, nor vice-versa.
We need to provide a semantics for attacks in order to judge whether such
syntactic refinements represent possible attacks.
To this end, we now formalise the semantics of attack trees by a proof 
theory.

\section{Proof Calculus}
\label{sec:procal}
A valid attack, intuitively, is one which is fully refined into fine-grained
attacks that are feasible in a model. The general model we provide is
a Kripke structure, i.e., a set of states and a generic state transition.
Thus, feasible steps in the model are single steps of the state transition.
We call them valid base attacks.
The composition of sequences of valid base attacks into and-attacks yields
again valid attacks if the base attacks line up with respect to the states
in the state transition. If there are different valid attacks for the same
attack goal starting from the same initial state set, these can be 
summarized in an or-attack.
\begin{ttbox}
{\bf fun} is_attack_tree :: [(\ttsigma :: state) attree] \ttfun bool  ("\ttvdash_") 
{\bf where} 
  att_base:  \ttvdash \ttcalN{s} = \ttforall x \ttin fst s. \ttexists y \ttin snd s. x  \ttrelI y  
| att_and: \ttvdash (As :: (\ttsigma::state attree list)) \ttattand{s} = 
           case As of
             [] \ttfun (fst s \ttsubseteq snd s)
           |  [a] \ttfun \ttvdash a \ttand attack a = s 
           |  a \# l \ttfun \ttvdash a \ttand fst(attack a) = fst s 
                         \ttand \ttvdash l \ttattand{\texttt{(snd(attack a),snd(s))}} 
| att_or: \ttvdash (As :: (\ttsigma::state attree list)) \ttattor{s} = 
          case As of 
             [] \ttfun (fst s \ttsubseteq snd s) 
          | [a] \ttfun \ttvdash a \ttand fst(attack a) \ttsupseteq fst s \ttand snd(attack a) \ttsubseteq snd s
          | a \# l \ttfun \ttvdash a \ttand fst(attack a) \ttsubseteq fst s \ttand snd(attack a) \ttsubseteq snd s
                       \ttand \ttvdash l \ttattor{\texttt{(fst s - fst(attack a),snd s)}}
\end{ttbox}
More precisely, the different cases of the validity predicate are distinguished
by pattern matching over the attack tree structure.
\begin{itemize}
\item A  base attack \texttt{\ttcalN{(s0,s1)}} is  valid if from all
states in the pre-state set \texttt{s0} we can get with a single step of the 
state transition relation to a state in the post-state set \texttt{s1}. Note,
that it is sufficient for a post-state to exist for each pre-state. After all,
we are aiming to validate attacks, that is, possible attack paths to some
state that fulfills the attack property.
\item An and-attack \texttt{As \ttattand{\texttt{(s0,s1)}}} is a valid attack
 if either of the following cases holds:
  \begin{itemize}
   \item empty attack sequence \texttt{As}: in this case 
       all pre-states in \texttt{s0} must already be attack states 
       in \texttt{s1}, i.e., \texttt{s0 \ttsubseteq\ s1};
   \item attack sequence \texttt{As} is singleton: in this case, the 
      singleton element attack \texttt{a} in \texttt{[a]}, 
      must be a valid attack and it must be an attack with pre-state 
      \texttt{s0} and post-state \texttt{s1};
   \item otherwise, \texttt{As} must be a list matching \texttt{a \# l} for
     some attack \texttt{a} and tail of attack list \texttt{l} such that
     \texttt{a} is a valid attack with pre-state identical to the overall 
     pre-state \texttt{s0} and the goal of the tail \texttt{l} is 
     \texttt{s1} the goal of the  overall attack. The pre-state of the
     attack represented by \texttt{l} is \texttt{snd(attack a)} since this is 
     the post-state set of the first step \texttt{a}.
     
   \end{itemize}
 \item An or-attack \texttt{As \ttattor{(s0,s1)}} is a valid attack 
  if either of the following cases holds: 
  \begin{itemize}
   \item the empty attack case is identical to the and-attack above: \texttt{s0 \ttsubseteq\ s1};
   \item attack sequence \texttt{As} is singleton: in this case, the 
      singleton element attack \texttt{a} 
      must be a valid attack and 
      its pre-state must include the overall attack pre-state set \texttt{s0} 
      (since \texttt{a} is singleton in the or) while the post-state of 
      \texttt{a} needs to be included in the global attack goal \texttt{s1};
   \item otherwise, \texttt{As} must be a list  \texttt{a \# l} for
     an attack \texttt{a} and a list \texttt{l} of alternative attacks.
     The pre-states can be just a subset of \texttt{s0} (since there are
     other attacks in \texttt{l} that can cover the rest) and the goal
     states \texttt{snd(attack a)} need to lie all in the overall goal
     state set \texttt{s1}. The other or-attacks in \texttt{l} need
     to cover only the pre-states \texttt{\texttt{fst s - fst(attack a)}}
     (where \texttt{-} is set difference) and have the same goal \texttt{snd s}.
   \end{itemize}
\end{itemize}

The proof calculus is thus completely described by one recursive function. 
This is a major
improvement to the inductive definition provided in the preliminary 
workshop paper \cite{kam:17a} that inspired this paper. Our notion 
of attack tree validity is more concise hence less prone to stating
inconsistent definitions and still allows to infer properties important
for proofs.  
The increase of consistency is because other important or useful algebraic 
properties can be derived from the recursive function definition. 
Note, that preliminary experiments on a proof calculus for attack 
trees in Isabelle \cite{kam:17a} used an inductive definition 
that had a larger number of rules than the three cases we have in our recursive function 
definition \texttt{is\_attack\_tree}. The earlier inductive definition integrated 
a fair number of properties as inductive
rules which are now proved from the three cases of \texttt{is\_attack\_tree}.

It might appear that Kripke semantics interprets conjunction as sequential 
(ordered) conjunction instead of parallel (unordered) conjunction. However,
this is not the case: the ordering of events or actions is implicit in the
states. Therefore, any kind of interleaving (or true parallelism) of state 
changing actions is possible. This is 
inserted as part of the application -- for example in the Infrastructures 
definition of the state transition in Section \ref{sec:iothc}. There the order 
of actions between states depends on the pre-states and post-states only.

Given the proof calculus, the notion of validity of an attack tree can 
be used to identify valid refinements already at a more abstract level. 
The notion \texttt{\ttrefV} denotes that the refinement of the attack tree
on the left side is to a valid attack tree on the right side.
\begin{ttbox}
  A \ttrefV A' \ttequiv  ( A \ttref A' \ttand  \ttvdash A')
\end{ttbox}
Taking this one step further, we can say that an abstract attack tree
is valid if there is a valid attack tree it refines to.
\begin{ttbox}
 \ttvdashV A  \ttequiv  (\ttexists A'. A \ttrefV A')
\end{ttbox}

\section{Correctness and Completeness of Attack Trees}
\label{sec:cor}
The novel contribution of this paper is to equip attack trees with 
a Kripke semantics. Thereby, a valid attack tree corresponds to an 
attack sequence. The following theorem provides this.
\begin{ttbox}
{\bf theorem} AT_EF: \ttvdash A :: (\ttsigma :: state) attree) \ttImp (I, s) = attack A 
 \ttImp Kripke \{s . \ttexists i \ttin I. i \ttrelI^* s \} I \ttvdash {\sf EF} s
\end{ttbox}
It is not only an academic exercise to prove this theorem. Since we use an embedding
of attack trees into Isabelle, this kind of proof about the embedded notions
of attack tree validity $\vdash$ and CTL formulas like {\sf EF} is possible.
At the same time, the established relationship between these notions can be applied
to case studies. Consequently, if we apply attack tree refinement to spell out
an abstract attack tree for attack \texttt{s} into a valid attack sequence, we can apply 
theorem \texttt{AT\_EF} and can immediately infer that {\sf EF} \texttt{s} holds.


Theorem \texttt{AT\_EF} also extends to validity of abstract attack trees.
That is, if an ``abstract'' attack tree \texttt{A} can be refined to a valid attack tree,
correctness in CTL given by \texttt{AT\_EF} applies also to the abstract tree 
as expressed in the following theorem.
\begin{ttbox}
{\bf theorem} ATV_EF: \ttvdashV A :: (\ttsigma :: state) attree) \ttImp (I, s) = attack A 
 \ttImp Kripke \{s . \ttexists i \ttin I. i \ttrelI^* s \} I \ttvdash {\sf EF} s
\end{ttbox}

The inverse direction of theorem \texttt{ATV\_EF} is a completeness
theorem.
\begin{ttbox} 
{\bf theorem} Completeness: I \ttneq \{\} \ttImp finite I \ttImp
 Kripke \{s . \ttexists i \ttin I. i \ttrelI^* s \} I \ttvdash {\sf EF} s 
 \ttImp \ttexists A :: (\ttsigma::state)attree. \ttvdashV A \ttand (I, s) = attack A 
\end{ttbox}


\section{Application to Infrastructures, Policies, and Actors}
\label{sec:iothc}
The Isabelle Infrastructure framework supports the representation of infrastructures 
as graphs with actors and policies attached to nodes. These infrastructures 
are the {\it states} of the Kripke structure. 

The transition between states is triggered by non-parametrized
actions \texttt{get}, \texttt{move}, \texttt{eval}, and \texttt{put} 
executed by actors. 
Actors are given by an abstract type \texttt{actor} and a function 
\texttt{Actor} that creates elements of that type from identities 
(of type \texttt{string}). Policies are given by pairs of predicates 
(conditions) and sets of (enabled) actions.
\begin{ttbox}
{\bf type}_{\bf{synonym}} policy = ((actor \ttfun bool) \tttimes action set)
\end{ttbox}
Actors are contained in an infrastructure graph.
\begin{ttbox}
{\bf datatype} igraph = Lgraph (location \tttimes location)set 
                           location \ttfun identity set
                           actor \ttfun (string set \tttimes string set)  
                           location \ttfun (string \tttimes acond)
\end{ttbox}
An \texttt{igraph} contains
a set of location pairs representing the 
topology of the infrastructure
as a graph of nodes and a list of actor identities associated to each node 
(location) in the graph.
Also an \texttt{igraph} associates actors to a pair of string sets by
a pair-valued function whose first range component is a set describing
the credentials in the possession of an actor and the second component
is a set defining the roles the actor can take on. More importantly in this 
context, an  \texttt{igraph} assigns locations to a pair of a string that defines
the state of the component and an element of type \texttt{acond}. This
type \texttt{acond} is defined as a set of labelled data representing a condition
on that data.
Corresponding projection functions for each of these components of an 
\texttt{igraph} are provided; they are named \texttt{gra} for the actual set of pairs of
locations, \texttt{agra} for the actor map, \texttt{cgra} for the credentials,
and \texttt{lgra} for the state of a location and the data at that location.

Infrastructures are given by the following datatype that contains
an infrastructure graph of type \texttt{igraph} and a policy 
given by a function that assigns local policies over a graph to
all locations of the graph. 
\begin{ttbox}
{\bf{datatype}} infrastructure = Infrastructure igraph 
           \;                              igraph \ttfun location \ttfun policy set
\end{ttbox}
There are projection functions \texttt{graphI} and \texttt{delta} when applied
to an infrastructure return the graph and the policy, respectively.
Policies specify the expected behaviour of actors of an infrastructure. 
They are defined by the \texttt{enables} predicate:
an actor \texttt{h} is enabled to perform an action \texttt{a} 
in infrastructure \texttt{I}, at location \texttt{l}
if there exists a pair \texttt{(p,e)} in the local policy of \texttt{l}
(\texttt{delta I l} projects to the local policy) such that the action 
\texttt{a} is a member of the action set \texttt{e} and the policy 
predicate \texttt{p} holds for actor \texttt{h}.
\begin{ttbox}
enables I l h a \ttequiv \ttexists (p,e) \ttin delta I l. a \ttin e \ttand p h
\end{ttbox} 
We now flesh out the abstract state transition introduced in Section \ref{sec:kripke}
by defining an inductive relation \texttt{\ttrel{n}} for state transition between 
infrastructures. This state transition relation is dependent on actions but also on
enabledness and the current state of the infrastructure. For illustration purposes
we consider the rule for \texttt{get\_data} only 
(see the complete source code \cite{kam:18smc} for full details of other rules)\footnote{We
deliberately omit here the DLM constraints for illustration purposes (see below)}.
\begin{ttbox}
get_data : G = graphI I \ttImp a \ttatI l \ttImp
           enables I l' (Actor a) get \ttImp
           I' = Infrastructure 
                 (Lgraph (gra G)(agra G)(cgra G)
                         ((lgra G)(l := (fst (lgra G l), 
                                   snd (lgra G l) \ttcup \{new\}))))
                         (delta I)
           \ttImp I \ttrel{n} I'
\end{ttbox}

\subsection{Application Example from IoT Healthcare}
\label{sec:app}
\begin{figure}[h]
\includegraphics[scale=.25]{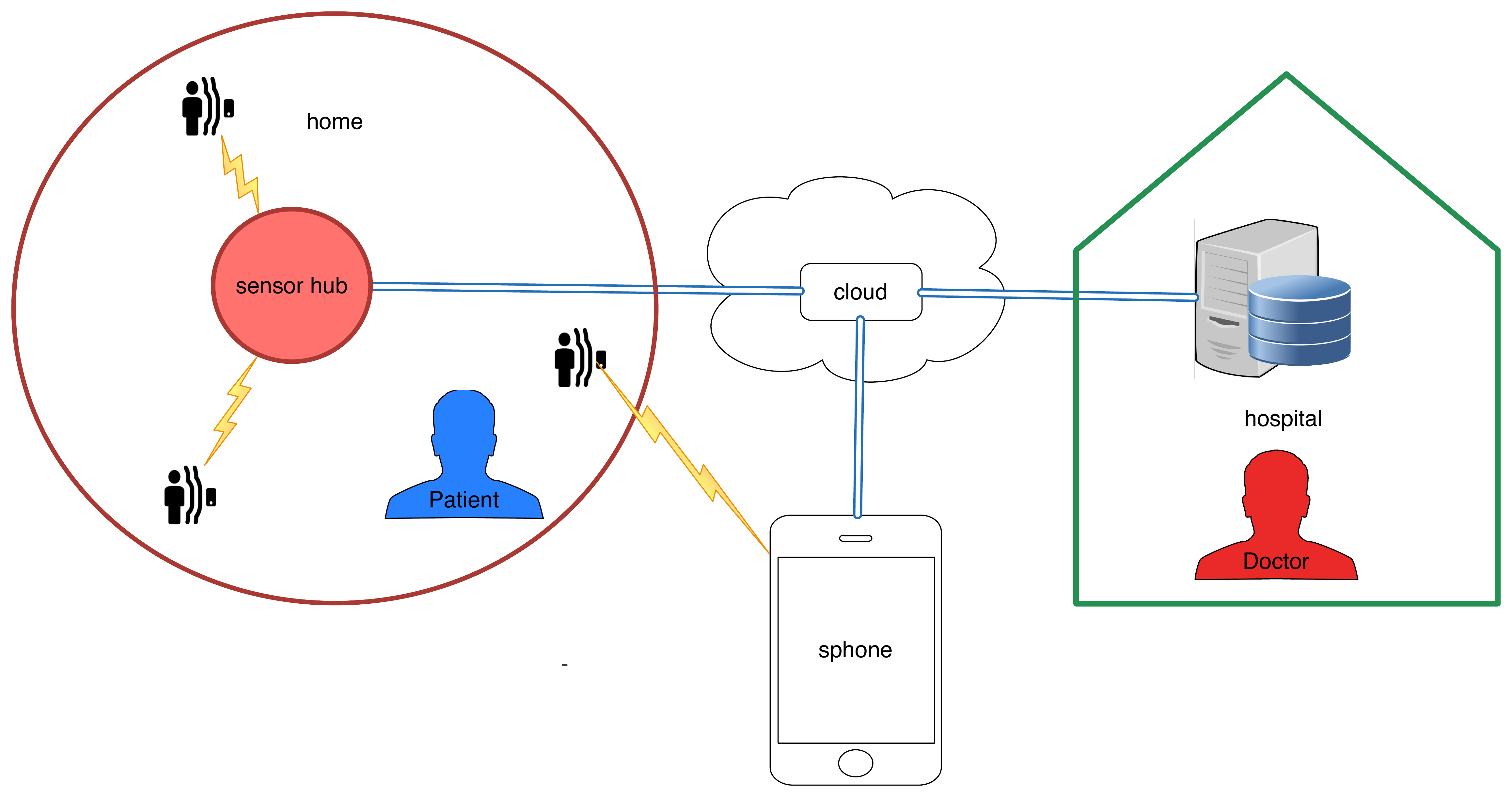}
\caption{IoT healthcare monitoring system for SUCCESS project}\label{fig:iot}
\end{figure}

%
The  example of an IoT healthcare systems is from the CHIST-ERA project SUCCESS \cite{suc:16}
on monitoring Alzheimer's patients. 
Figure \ref{fig:iot} illustrates the system architecture where data collected by sensors 
in the home or via a smart phone helps monitoring bio markers of the patient. The data 
collection is in a cloud based server to enable hospitals (or scientific institutions) 
to access the data which is controlled via the smart phone.
We show the encoding of the \texttt{igraph} for this system architecture in the Infrastructure model.
\begin{ttbox}
ex_graph \ttequiv Lgraph \{(home, cloud), (sphone, cloud), (cloud,hospital)\}
                   (\ttlam x. if x = home then \{''Patient''\} 
                         else (if x = hospital then \{''Doctor''\} else \{\})) 
                         ex_creds ex_locs
\end{ttbox}
The identities \texttt{Patient} and \texttt{Doctor} represent patients
and their doctors; double quotes \texttt{''s''} indicate strings 
in Isabelle/HOL.
The global policy is `only the patient and the doctor can access the data in the cloud':
\begin{ttbox}
{\bf fixes} global_policy::[infrastructure, identity] \ttfun bool
{\bf defines}  global_policy I a \ttequiv  a \ttnin gdpr\_actors \ttimp 
                  \ttneg(enables I cloud (Actor a) get)
\end{ttbox}

Local policies are represented as a function over an \texttt{igraph G} that 
additionally assigns each location of a scenario to its local policy
given as a pair of requirements to an actor (first element of the pair)
in order to grant him actions in the location (second element of the pair).
The predicate $\ttatI$ checks whether an actor is at a given location in 
the graph $G$.
\begin{ttbox}
local_policies G \ttequiv 
(\ttlam x. if x = home then \{(\ttlam y. True, \{put,get,move,eval\})\}
 else (if x = sphone then 
  \{((\ttlam y. has G (y, ''PIN'')), \{put,get,move,eval\})\} 
    else (if x = cloud then \{(\ttlam y. True, \{put,get,move,eval\})\}
          else (if x = hospital then
                \{((\ttlam y. (\ttexists n. (n \ttatI hospital) \ttand Actor n = y \ttand 
                 has G (y, ''skey''))), \{put,get,move,eval\})\} else \{\}))))
\end{ttbox}

\subsection{Using Attack Tree Calculus}
\label{sec:hcat}
%
%

Since we consider a predicate transformer semantics, we use
sets of states to represent properties. For example, the attack property
is given by the following set \texttt{sgdpr}.
\begin{ttbox}
  sgdpr \ttequiv \{x. \ttneg (global_policy x ''Eve'')\}
\end{ttbox}
%
%
The attack we are interested in is to see whether for the scenario
\begin{ttbox}
 gdpr\_scenario \ttequiv  Infrastructure ex_graph local_policies 
\end{ttbox}
from the initial state \texttt{Igdpr \ttequiv \{gdpr\_scenario\}},
the critical state \texttt{sgdpr} can be reached,
i.e., is there a valid attack \texttt{(Igdpr,sgdpr)}?

For the Kripke structure
\begin{ttbox}
 gdpr_Kripke \ttequiv Kripke \{ I. gdpr_scenario \ttrelIstar I \} Igdpr
\end{ttbox}
we first derive a valid and-attack using the attack tree proof calculus.
\begin{ttbox}
\ttvdash [\ttcalN{(Igdpr,GDPR)},\ttcalN{(GDPR,sgdpr)}]\ttattand{\texttt{(Igdpr,sgdpr)}}
\end{ttbox}
The set \texttt{GDPR} is an intermediate state where \texttt{Eve} accesses the cloud.
We can then simply apply the Correctness theorem \texttt{AT\_EF} to 
immediately prove the following CTL statement.
\begin{ttbox}
 gdpr_Kripke \ttvdash {\sf EF} sgdpr
\end{ttbox}
This application of the meta-theorem of Correctness of attack trees
saves us proving the CTL formula tediously by exploring the state space.


\section{Data Protection by Design for GDPR compliance}
\label{sec:gdpr}

\subsection{General Data Protection Regulation (GDPR)}
From 26th May 2018, the GDPR will become mandatory within the European Union and hence
also for any supplier of IT products. Breaches of the regulation will be fined with 
penalties of 20 Million EUR.
 For this paper, we use the final proposal \cite{gdpr:18}
as our source. Despite the relatively large size of
the document of 209 pages, the technically relevant portion for us is only about 30 
pages (Pages 81--111, Chapters I to Chapter III, Section 3).
In summary, Chapter III specifies that the controller must give the data subject 
{\it read access} (1) to any information, communications, and ``meta-data'' of the 
data, e.g., retention time and purpose. In addition, the system must enable 
{\it deletion of data} (2) and restriction of processing. 

An invariant condition for data processing resulting from these Articles 
is that the system {\it functions} must {\it preserve} any of the access 
rights of personal data (3).

\subsection{Security and Privacy by Labeling Data}
\label{sec:label}
The Decentralised Label Model (DLM) \cite{ml:98} introduced the idea to
label data by owners and readers. We pick up this idea and formalize 
a new type to encode the owner and the set of readers.
\begin{ttbox}
{\bf type\_synonym} dlm = actor \tttimes actor set
\end{ttbox}
Labelled data is then just given by the type \texttt{dlm \tttimes\ data}
where \texttt{data} can be any data type. Additional meta-data, like retention
time and purpose, can be encoded as part of this type \texttt{data}. We omit
these detail here for conciseness of the exposition.

Using labeled data, we can now express the essence of Article 4
Paragraph (1): 'personal data' means any information relating to an 
identified or identifiable natural person ('data subject').
Since we have a more constructive system view, we express this by
defining the owner of a data item \texttt{d} of type \texttt{dlm}
as the actor that is the first element in the pair that is the first 
of the pair \texttt{d}.
Then, we use this function to express the predicate ``owns''. 
\begin{ttbox}
{\bf definition} owner :: dlm \tttimes data \ttfun actor 
{\bf where} owner d \ttequiv fst(fst d)

{\bf definition} owns :: [igraph, location, actor, dlm \tttimes data] \ttfun bool    
{\bf where} owns G l a d \ttequiv owner d = a
\end{ttbox}    
The introduction of a similar function for readers projecting the second element
of a \texttt{dlm} label 
\begin{ttbox}
{\bf definition} readers :: dlm \tttimes data \ttfun actor set
{\bf where} readers d \ttequiv snd (fst d)
\end{ttbox}
enables specifying 
whether an actor may access a data item.
\begin{ttbox}    
{\bf definition} has_access :: [igraph, location, actor, dlm \tttimes data] \ttfun bool    
{\bf where} has_access G l a d \ttequiv owns G l a d \ttor a \ttin readers d
\end{ttbox}

For our example of an IoT health care monitoring system, the data and its privacy 
access control definition is given by 
the parameter \texttt{ex\_locs} specifying that the data 42, for example 
some bio marker's value, is owned by the patient and can be read by the doctor
(\texttt{''free''} is the state of the cloud component).
\begin{ttbox}
ex_locs \ttequiv (\ttlam x. if x = cloud 
           then (''free'', \{((Actor ''Patient'',\{Actor ''Doctor''\}),42)\}) 
           else ('''',\{\}))
\end{ttbox}

\subsection{Privacy Preserving Functions}
\label{sec:ppfun}
The labels of data must not be changed by processing: we have identified this
finally as an invariant (3) resulting from the GDPR in Section \ref{sec:gdpr}.
This invariant can be formalized in our Isabelle model by a type definition 
of functions on labeled data that preserve their labels.
\begin{ttbox}
{\bf typedef} label_fun = \{f :: dlm\tttimes{data} \ttfun dlm\tttimes{data}. \ttforall x. fst x = fst (f x)\}  
\end{ttbox}
We also define an additional function application operator \texttt{\ttupdownarrow}
on this new type. Then we can use this restricted function type to implicitly
specify that only functions preserving labels may be applied in the definition
of the system behaviour in the 
state transition rule for action \texttt{eval} (see \cite{kam:18smc}).

\subsection{Policy Enforcement}
We can now use the labeled data to encode the privacy constraints of the GDPR
in the rules. For example, the \texttt{get\_data} rule has now labelled data
\texttt{((Actor a', as), n)} and used the labeling in the precondition to guarantee
that only entitled users can get data.
\begin{ttbox}
get_data : G = graphI I \ttImp a \ttatI l \ttImp enables I l' (Actor a) get \ttImp
           ((Actor a', as), n) \ttin snd (lgra G l') \ttImp Actor a \ttin as \ttImp
           I' = Infrastructure 
                 (Lgraph (gra G)(agra G)(cgra G)
                         ((lgra G)(l := (fst (lgra G l), 
                             snd (lgra G l) \ttcup \{((Actor a', as), new)\}))))
                         (delta I)
           \ttImp I \ttrel{n} I'
\end{ttbox}
Using the formal model of infrastructures, we can now 
prove privacy by design for GDPR compliance of the specified system.
We can show how the properties relating to data ownership, processing and deletion
can be formally captured using Kripke structures and CTL and the Infrastructure
framework. As an example, consider the preservation of data ownership.

\subsubsection{Processing preserves privacy}
\label{sec:pres}
We can prove that processing preserves ownership as defined
in the initial state for all paths globally (\texttt{\sf AG}) 
within the Kripke structure and in all locations of the graph.
\begin{ttbox}  
{\bf theorem} GDPR_three: h \ttin gdpr_actors \ttImp l \ttin gdpr_locations \ttImp 
  owns (Igraph gdpr_scenario) l (Actor h) d \ttImp
  gdpr_Kripke \ttvdash 
    {\sf AG} \{x. \ttforall l \ttin gdpr_locations. owns (Igraph x) l (Actor h) d \}  
\end{ttbox}
Note, that it would not be possible to express 
this property in Modelcheckers since they only allow propositional logic 
within states.
This generalisation is only possible since we use Higher Order 
Logic. 

The proved meta-theory for attack trees can be applied to facilitate
the proof.
The contraposition of the Completeness property grants that if there
is no attack on \texttt{(I,$\neg f$)}, then {\sf (EF $\neg f$)}
does not hold in the Kripke structure. 
This yields the theorem since the  {\sf AG $f$} statement corresponds 
to $\neg${\sf (EF $\neg f$)}.

\section{Conclusions}
\label{sec:concl}
\subsection{Summary and Discussion}
In this paper, we have presented a proof theory for attack trees
in Isabelle's Higher Order Logic (HOL).
We have shown the incremental and generic structure of this framework,
presented correctness and completeness results equating valid attacks 
to {\sf EF} $s$ formulas.  
The proof theory has been illustrated on an IoT healthcare infrastructure where 
the meta-theorem of completeness could be directly applied to infer the existence 
of an attack tree from CTL. 
The practical relevance has been demonstrated on GDPR compliance verification.

The use of an expressive proof assistant like Isabelle comes at a cost:
all proofs have to be done in interaction with the user. On the application
side this seems to be much inferior to using fully automated techniques like
SMT solvers or model checkers that have been employed heavily in security
applications including attack tree analysis, for example \cite{anp:16}. 
However, and this is the 
advantage that we claim for our approach, none of these related works can
claim the same level of consistency of the used concepts and formalisms.
Since HOL is so expressive, we combine meta-theoretic reasoning with
application level verification in one logical theory. Thereby, we can
prove meta-theorems, like the presented correctness results, and 
simultaneously use the formalised concepts of attack trees and temporal 
logic on applications. This can be done in a nicely generic and structured
way using type classes and instantiation in Isabelle.
Powerful concepts like recursive function in HOL and simplification and
other proof tactics in Isabelle furthermore facilitate the application.

\subsection{Related Work}
There are excellent foundations available based on graph theory \cite{kps:14}. 
They provide a very good understanding of the formalism, various extensions 
(like attack-defense trees \cite{kmrs:14}) and differentiations of the 
operators (like sequential conjunction (SAND) versus parallel conjunction 
 \cite{jkmrt:15}) and are amply documented in the literature.
These theories for attack trees provide a thorough foundation for the 
formalism and its semantics. The main problem that adds complexity to the 
semantical models is the abstractness of the descriptions in the nodes. This 
leads to a variety of approaches to the semantics, e.g. propositional 
semantics, multiset semantics, and equational semantics for ADtrees 
\cite{kmrs:14}. The theoretical foundations allow comparison of different 
semantics, and provide a theoretical framework to develop evaluation 
algorithms for the quantification of attacks.


More practically oriented formalisations, e.g. \cite{anp:16}, focus on an 
action based-approach where the attack goals are represented as labels of 
attack tree nodes which are actions that an attacker has to execute to 
arrive at the goal.

A notable exception that uses, like our approach, a state based semantics for
attack trees is the recent work \cite{apk:17}. However, this work is aiming at 
assisted generation of attack trees from system models.
The tool ATSyRA supports this process. The paper 
\cite{apk:17} focuses on describing a precise 
semantics of attack tree in terms of transition systems using ``under-match'', 
``over-match'', and ``match'' to arrive at a notion of correctness.
In comparison, we use additionally CTL logic to describe the correctness
relation precisely. Also we use a fully formalised and proved Isabelle model.

Surprisingly, the use of an automated proof assistant, like Isabelle, has not 
been considered before despite its potential of providing a theory and analysis of 
attacks simultaneously. The essential attack tree mechanism of disjunction 
and conjunction in tree refinement is relatively simple.
The complexity in the theories is caused by the attempt to incorporate 
semantics to the attack nodes and relate the trees to actual scenarios. This 
is why we consider the formalisation of a foundation of attack trees in the 
interactive prover Isabelle since it supports logical modeling and 
definitions of datatypes very akin to algebraic specification
but directly supported by semi-automated analysis and proof tools.

The workshop paper \cite{kam:17a} has inspired the present work but is vastly superseded by it.
The novelties are:
\begin{itemize}
\item Attack trees have a formal state based semantics that is formalised in the framework.
\item Correctness and completeness are proved with respect to the formal semantics.
\item The Isabelle framework is generic using type classes, that is, works for any state model.
      Infrastructures with actors and policies are an instantiation.
\item The semantics, correctness and completeness theorems facilitate application verification.
\end{itemize}

\subsection{Outlook}
The high level analysis of infrastructures including humans and 
high level policies specifying the (desired) behaviour of systems and humans
alike is a great challenge to formal methods.
The approach taken in the Isabelle insider framework and prolonged in
the attack tree proof theory presented in this paper follows an approach
akin to formal software engineering but integrating high level ideas
from sociology and tapping into results from psychology \cite{kp:16}.
We believe that the current presented attack tree proof calculus is
an important step towards complementing this endeavour by providing 
a generic yet rigorous frame to allow the use of the established and
easily understandable attack tree concept to be used within the wider
framework. This is particularly relevant for transparency of security
and privacy, one of the goals of the SUCCESS project \cite{suc:16}
supporting this research.

The genericity of the attack tree formalisation presented
here will 
permit the extension of the CTL temporal logic 
to specialised modal logics like the actor logic ATL \cite{ahk:02}
or by adding probabilities. The extension by defences to attack-defence 
trees, 
implies the change of the Kripke model. Since Kripke models are first 
class citizen of the logic, 
we can integrate Kripke model transformations into the framework to allow 
defences to detected attacks by model refinement.
This can be considered as yet another avenue for future research.



\end{document}